\documentclass[11pt]{article}

\usepackage{graphicx}
\usepackage[T1]{fontenc}
\usepackage{booktabs}
\usepackage{tabularx}
\usepackage{array}
\usepackage{cite}
\usepackage{url}
\usepackage{microtype}
\usepackage{float}
\usepackage{placeins}
\usepackage{needspace}
\usepackage{longtable}
\usepackage[margin=1in]{geometry}
\usepackage[hidelinks]{hyperref}

\newcolumntype{Y}{>{\raggedright\arraybackslash}X}

\title{Beyond Network Topology: Biological Evidence Integration and Reproducible Benchmarking for Protein Complex Detection}
\hypersetup{pdftitle={Beyond Network Topology: Biological Evidence Integration and Reproducible Benchmarking for Protein Complex Detection}, pdfauthor={Sima Soltani, Mehrdad Jalali, Yahya Forghani, and Reza Sheibani}}
\author{Sima Soltani\textsuperscript{1}, Mehrdad Jalali\textsuperscript{2,*}, Yahya Forghani\textsuperscript{1}, and Reza Sheibani\textsuperscript{1}\\[0.5em]
\small \textsuperscript{1}Department of Computer Engineering, Ma.C., Islamic Azad University, Mashhad, Iran\\
\small \textsuperscript{2}Applied Data Science and Artificial Intelligence, SRH University Heidelberg, Heidelberg, Germany\\
\small \textsuperscript{*}Corresponding author: \texttt{mehrdad.jalali@srh.de}\\
\small Running title: Evidence-Aware Protein Complex Detection}
\date{}

\begin{document}
\maketitle

\begin{abstract}
Protein complexes are molecular assemblies that coordinate cellular regulation, signaling, metabolism, and disease-relevant protein function. Detecting such assemblies from protein-protein interaction (PPI) networks remains difficult because network topology is an incomplete abstraction: an edge may reflect direct binding, functional association, co-complex evidence, co-expression, co-localization, or a computationally predicted interaction. This focused critical methodological review examines how biological evidence can improve protein-complex detection beyond dense-subgraph discovery. The synthesis covers Gene Ontology (GO), expression, localization, domains and motifs, sequence and structure, interface evidence, temporal context, RNA or regulatory evidence, and representation learning, while retaining classical graph-clustering methods as historical baselines. Interpretable evidence-aware graph methods currently provide a strong balance between biological plausibility and reproducibility, whereas structure-aware, temporal, heterogeneous, and hypergraph models offer greater biological realism but still need independent benchmarking. Reported F-measures cannot be ranked across incompatible PPI releases, reference sets, matching thresholds, preprocessing pipelines, and metric implementations. We argue that progress requires fixed dataset versions, explicit GO-circularity controls, overlap-aware metrics, uncertainty estimates, and executable software packages. Reliable protein-complex detection therefore requires connecting graph-based predictions to molecular structure, interaction mechanisms, cellular context, and functional assembly.
\end{abstract}

\noindent\textbf{Keywords:} Protein complex detection; protein-protein interaction networks; structural bioinformatics; biological evidence integration; protein interfaces; multi-omics integration; graph learning; reproducible benchmarking

\section{Introduction}
Protein complexes are physical or condition-dependent molecular assemblies whose subunits act together in transcription, chromatin regulation, signaling, cell-cycle control, metabolism, protein folding, and proteostasis \cite{r1,r2,r3,r4,r5,r6,r54}. Their detection from PPI networks is therefore not only a graph-mining problem but also a structure-function inference problem. A predicted group of proteins becomes biologically convincing only when its topology is compatible with functional coherence, cellular compartment, interaction mechanism, and the biological context in which the assembly can form.

\begin{figure}[!t]
\centering
\includegraphics[width=\textwidth]{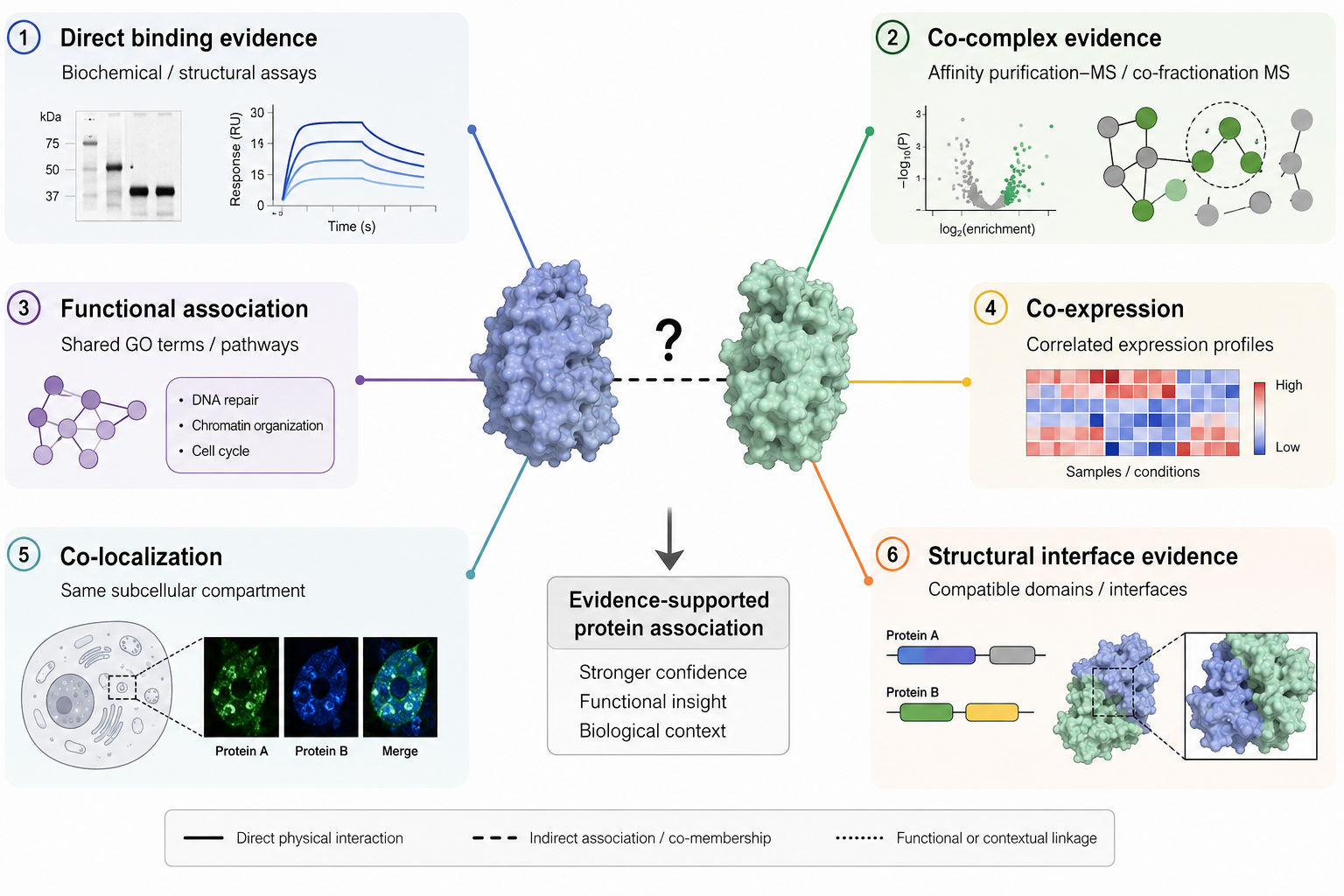}
\caption{Evidence-based protein-association schematic for protein-complex detection. The figure distinguishes direct binding evidence, co-complex evidence, functional association, co-expression, co-localization, and structural interface evidence, illustrating how heterogeneous association signals can support or qualify candidate molecular assemblies. \emph{Source note: conceptual illustration prepared for this review.}}
\label{fig:ppi-concept}
\end{figure}

A PPI network is a heterogeneous abstraction of molecular interaction evidence. A graph edge may represent direct physical binding, membership in the same experimentally purified complex, a curated functional association, co-expression, co-localization, or a predicted interaction \cite{r1,r2,r35,r41}. These categories should not be treated as equivalent. Direct binding implies compatible interfaces, whereas co-complex membership may be mediated by additional subunits; co-expression can indicate shared regulation without physical contact; and co-localization is necessary but not sufficient for assembly. Figure~\ref{fig:ppi-concept} summarizes an evidence-based protein-association perspective for complex detection.

Classical methods such as MCODE, MCL, CFinder, ClusterOne, and RNSC established that protein complexes can often be approximated by dense or overlapping graph regions \cite{r7,r8,r9,r10,r11}. These baselines remain important for reproducible comparison, but their assumptions are biologically limited. True assemblies can be sparse under a particular PPI release, transient under specific conditions, shared through scaffold proteins, or absent from a static graph because the relevant tissue, time point, or structural interface was not measured \cite{r12,r25,r47,r54}. Modern methods therefore increasingly integrate topology with GO annotations, expression profiles, subcellular localization, domains, motifs, sequence or structural features, temporal information, RNA-protein regulation, and learned representations \cite{r15,r16,r18,r19,r23,r28,r29,r30,r31,r33,r34,r48,r51,r52,r56,r57,r58,r60}.

This review makes six contributions. First, it organizes biological evidence according to its molecular meaning rather than only by algorithmic convenience. Second, it analyzes where evidence enters computational pipelines: edge filtering, edge reweighting, seed selection, objective functions, representation learning, and candidate validation. Third, it distinguishes direct complex-detection methods from adjacent PPI, interface, motif-mapping, and structure-prediction approaches. Fourth, it evaluates biological realism, molecular interpretability, overlap support, human transferability, code availability, and reproducibility across method families. Fifth, it identifies GO circularity, benchmark leakage, and source-specific performance reporting as major threats to fair comparison. Sixth, it proposes reporting standards and research priorities for structure-aware, uncertainty-aware, and reproducible protein-complex prediction.

\subsection{Scope and distinction from previous reviews}
Previous reviews have usefully summarized protein-complex prediction, graph clustering, community detection, algorithmic scalability, context-specific complexes, and general PPI-network analysis \cite{r3,r4,r5,r6,r54,r61,r62,r63}. A recent review in \emph{PROTEINS} provides a comprehensive analysis of efficient and scalable graph-clustering algorithms for protein-complex identification \cite{r61}. The present review addresses a complementary question: how functional, spatial, temporal, regulatory, sequence, domain, motif, interface, and structural evidence can transform graph communities into biologically credible molecular assemblies, and how such methods should be evaluated reproducibly. Table~\ref{tab:previous-reviews} positions the present review relative to earlier surveys without implying that those reviews were incomplete or incorrect.

\begin{table}[!t]
\caption{Positioning Relative to Previous Reviews}
\label{tab:previous-reviews}
\centering
\footnotesize
\begin{tabularx}{\textwidth}{p{0.18\textwidth}YYYY}
\toprule
Review & Primary emphasis & Biological evidence & Structural/interface focus & Reproducibility emphasis \\
\midrule
Srihari and Leong \cite{r4} & Computational prediction survey & Discussed as part of method landscape & Limited relative to graph methods & Limited by period \\
Srihari et al. \cite{r5} & Function, dynamics, and complex prediction & Strong biological framing & General mechanistic context & Not a reporting-checklist focus \\
Li et al. \cite{r6} & Computational approaches from PPI networks & Algorithmic and data context & Limited by available methods & Limited by period \\
Zahiri et al. \cite{r3} & Broad protein-complex prediction survey & Covers multiple data sources & General rather than central & Survey-level comparison \\
Omranian et al. \cite{r54} & Present state and challenges & Highlights challenges and data limits & Discussed as future direction & Stronger challenge framing \\
Csik{\'a}sz-Nagy et al. \cite{r62} & Context-specific protein-complex prediction & Central to context specificity & Explicit structural-biology bridge & Future-oriented discussion \\
Nie et al. \cite{r63} & Context-specific PPI-network construction & Omics and dynamic context & Structural and biophysical context & Validation challenges \\
Patra and Sahoo \cite{r61} & Efficient scalable graph clustering & Secondary to algorithmic taxonomy & Limited/general & Algorithm comparison and scalability \\
Present review & Evidence-aware molecular-assembly inference & Central organizing axis & Domains, motifs, interfaces, sequence, and structure & Benchmark harmonization and minimum reporting checklist \\
\bottomrule
\end{tabularx}
\end{table}

\section{Biological Interpretation of Protein-Complex Detection}
\subsection{Protein complexes as molecular assemblies}
A protein complex is more than a graph community. It is an assembly whose subunits are compatible in molecular function, physical interface, cellular compartment, abundance, and timing. Stable complexes often contain recurrent cores and accessory subunits, whereas transient complexes may assemble only under particular regulatory or environmental states. This distinction matters because algorithms that force every protein into a single dense cluster can miss biologically valid multiple membership or condition-specific assemblies \cite{r9,r10,r12,r20}.

\subsection{What a PPI edge represents}
PPI edges are heterogeneous. Experimental sources may report direct physical interactions, co-purification, co-fractionation, or literature-curated relationships; databases such as DIP, HPRD, STRING, MIPS, CYC2008, PCDq, and CORUM differ in organism scope, evidence type, confidence scoring, curation age, and reference-complex definitions \cite{r35,r36,r37,r38,r39,r40,r41,r43}. STRING edges require particular care because they can include functional associations as well as physical evidence \cite{r41}. A method that treats every edge as a direct binding event risks confusing pathway membership or shared annotation with molecular assembly.

\subsection{Why topology alone is insufficient}
\begin{figure}[!t]
\centering
\includegraphics[width=\textwidth]{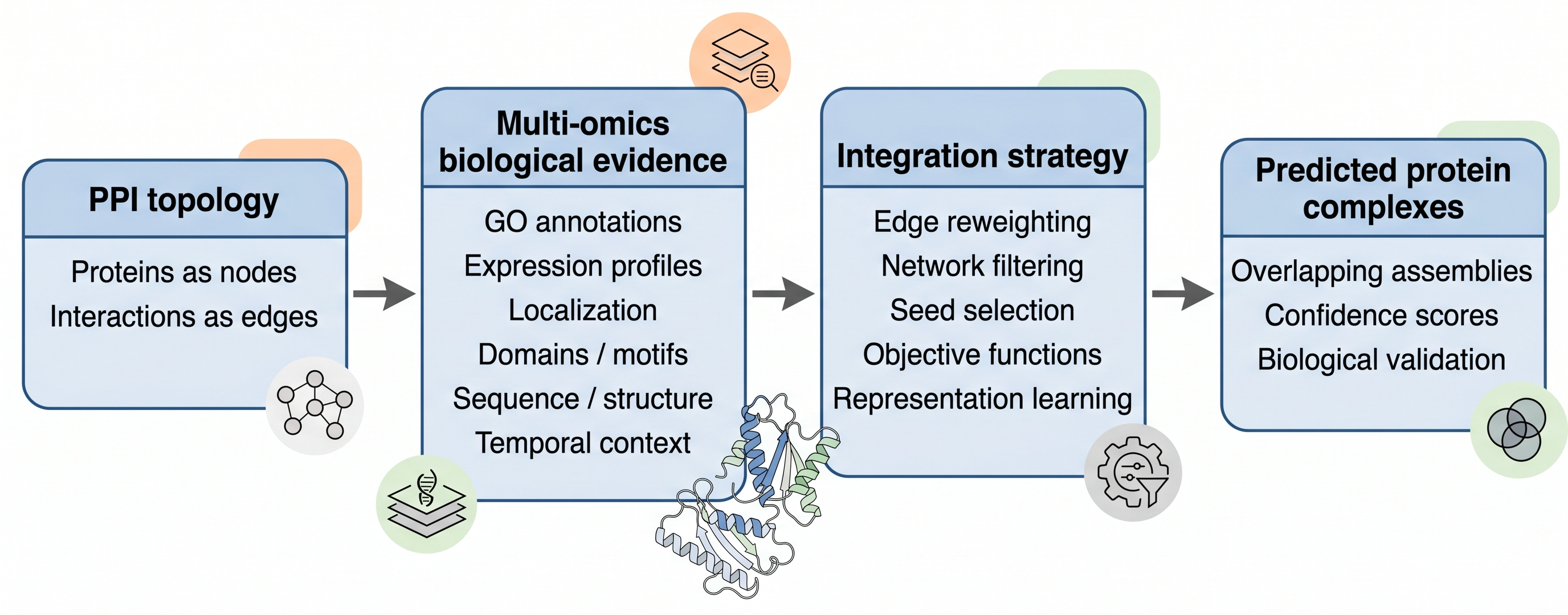}
\caption{From PPI evidence to candidate molecular assemblies. PPI topology provides the initial network structure, multi-omics biological evidence qualifies the molecular and cellular plausibility of candidate associations, and integration strategies convert heterogeneous evidence into predicted complexes with overlap, confidence, and validation context. \emph{Source note: conceptual workflow prepared for this review.}}
\label{fig:overview}
\end{figure}

Dense-subgraph discovery remains useful, but it is not sufficient. False-positive edges can create artificial clusters; missing edges can make real complexes appear sparse; hubs can connect unrelated assemblies; and static networks merge interactions that occur in different compartments, cell states, or time points. Topology should therefore define the search space, not the full biological interpretation. Figure~\ref{fig:ppi-concept} illustrates how heterogeneous biological and structural evidence can support or qualify candidate protein associations. Figure~\ref{fig:overview} summarizes the progression from heterogeneous PPI evidence to integrated, validated candidate molecular assemblies.

\section{Review Scope and Methodology}
This manuscript is a focused methodological review and targeted literature synthesis, not a PRISMA systematic review or meta-analysis. The objective is to explain methodological transitions and biological evidence integration rather than estimate pooled effect sizes. No PRISMA flow diagram or screening counts are reported because the original project did not generate auditable duplicate-removal logs, database-specific screening records, or exclusion statistics.

The search framing combined protein-complex terminology, PPI-network terminology, algorithmic terms, and biological-evidence terms. The following conceptual Boolean query was adapted to the syntax of each database: (``protein complex detection'' OR ``protein complex prediction'' OR ``protein complex identification'') AND (``protein-protein interaction'' OR PPI OR interactome) AND (clustering OR ``graph clustering'' OR ``network embedding'' OR hypergraph OR heterogeneous OR temporal OR dynamic OR ``Gene Ontology'' OR expression OR localization OR domain OR motif OR interface OR structure OR ``multi-omics''). Searches and metadata checks used PubMed, publisher pages, IEEE/ACM metadata, Crossref-indexed records where available, arXiv for preprint status, and targeted web searches for high-risk 2024-2026 references. Reference lists of included papers were backward-searched, and key recent papers were forward-searched through publisher or citation metadata where available. Table~\ref{tab:criteria} defines the inclusion, exclusion, and extraction framework used for this focused synthesis.

\begin{table}[!t]
\caption{Inclusion, Exclusion, and Extraction Framework}
\label{tab:criteria}
\centering
\footnotesize
\begin{tabularx}{\textwidth}{p{0.22\textwidth}YY}
\toprule
Category & Included or extracted & Excluded or treated as adjacent \\
\midrule
Task & Direct prediction or identification of protein complexes from PPI networks & Pairwise PPI prediction alone; essential-protein prediction; generic clustering without complex evaluation \\
Evidence & Topology plus GO, expression, localization, domain/motif, sequence/structure, temporal, regulatory, or learned evidence & Evidence claims not described sufficiently in the source \\
Method role & Classical baselines, integrative methods, dynamic/heterogeneous/hypergraph methods, and selected emerging contexts & Adjacent structure, interface, or PPI-prediction tools as direct benchmark competitors \\
Evaluation & Recognized or clearly described reference complexes, metrics, and source-specific protocols & Untraceable performance values or incomplete benchmark descriptions \\
Extraction fields & Method, year, evidence type, organism, benchmark context, overlap support, direct/adjacent status, interpretability, code availability, and reproducibility limitations & Fabricated counts, inferred DOI data, or unsupported performance values \\
\bottomrule
\end{tabularx}
\end{table}

Methods were selected to represent major transitions: topology baselines, pre-2018 integration, GO and expression integration, localization-aware methods, supervised and ensemble learning, network embeddings, evolutionary optimization, hypergraph learning, dynamic heterogeneous networks, RNA-protein regulatory integration, and adjacent structure/interface context. This scoping logic explains the representative set. The review does not claim exhaustive coverage of every PPI community-detection variant.

\section{Biological Evidence for Complex Detection}
\subsection{GO and functional coherence}
GO semantic similarity is useful because subunits of a complex often share biological-process, molecular-function, or cellular-component annotations \cite{r15,r16}. Methods use GO for edge weighting, seed expansion, candidate scoring, supervised features, or evolutionary objectives \cite{r17,r19,r21,r22,r23,r24,r26,r27,r51}. However, GO can overfavor well-annotated proteins and underrepresent sparsely annotated proteins. GO circularity can arise when GO is used both as an input feature and as part of interpretation, benchmark construction, or validation. Future studies should report GO releases, perform GO ablation or permutation controls, use time-stamped annotation policies where possible, and evaluate performance on sparsely annotated proteins.

\subsection{Expression and condition-specific assembly}
Expression data can help distinguish condition-specific assemblies from static network neighborhoods. Edge-weighted MCL, IPC-RPIN, NRAGE-WPN, DHPRL, MComplex, and STRPCI illustrate how expression or temporal expression can reweight networks or support dynamic representations \cite{r18,r19,r25,r47,r56,r57,r58}. The biological limitation is that transcript co-expression is an indirect proxy for protein abundance and assembly. Expression evidence is strongest when the condition, tissue, or time course matches the PPI context.

\subsection{Subcellular localization}
Co-localization is a basic biological constraint: proteins in incompatible compartments are less likely to form a stable complex. Localization-aware methods can reduce false-positive edges and support candidate validation \cite{r23,r28,r51,r58}. The limitation is that localization annotations can be coarse, dynamic, isoform-specific, or condition dependent. Localization should therefore constrain plausibility rather than act as absolute proof of co-complex membership.

\subsection{Domains and short linear motifs}
Domains and short linear motifs provide mechanistic evidence that graph topology alone cannot supply. Domain-domain and domain-motif interactions can indicate whether two proteins have compatible binding modules, whether a predicted edge is more likely to represent direct contact, and whether a candidate complex boundary is mechanistically plausible \cite{r29,r30,r31}. Motif evidence is especially relevant for transient interactions and intrinsically disordered regions, where short recognition elements can mediate regulated assembly. Its main risk is overprediction: short motifs are common, degenerate, and context dependent, so motif matches require localization, expression, conservation, or structural support before they are interpreted as complex-level evidence.

\subsection{Structural and interface evidence}
Structural evidence can help distinguish direct interaction from indirect co-complex membership. Experimentally resolved interfaces, predicted complex models, interface-confidence measures, contact compatibility, and predicted aligned-error patterns can support edge filtering, candidate-complex validation, boundary refinement, and prioritization for experimental follow-up. PPI-ID illustrates an adjacent domain- and SLiM-supported PPI/interface-mapping tool that maps compatible domains and motifs onto predicted three-dimensional structures and filters contacts by distance; it should inform complex-detection pipelines but should not be treated as a direct complex-detection benchmark competitor \cite{r48}. HyperGraphComplex integrates sequence and PPI evidence directly for complex identification \cite{r52}, whereas RF2-PPI is a structure-informed PPI-prediction advance for the human proteome, not a complex-detection algorithm \cite{r49}. Structure prediction and protein language models are therefore best treated as evidence sources or feature generators unless a study explicitly validates complete complex detection. Large, flexible, transient, or stoichiometrically variable assemblies remain difficult for structure-only interpretation because pairwise interface plausibility does not guarantee simultaneous co-complex membership.

\subsection{Experimental validation of predicted complexes}
Computational predictions ultimately require experimental validation at the appropriate evidence level. Affinity purification--mass spectrometry and co-fractionation mass spectrometry can support co-complex membership, but they may not resolve direct interfaces or transient assemblies \cite{r2,r63}. Cross-linking mass spectrometry, cryo-electron microscopy, crystallography, proximity labeling, and mutational or interface-disruption studies provide stronger mechanistic evidence when they localize contacts, preserve cellular context, or perturb a predicted interface. A validation plan should therefore state whether the goal is to confirm physical binding, co-complex association, spatial proximity, functional cooperation, or condition-specific assembly.

\subsection{Temporal, spatial, and regulatory evidence}
Temporal, spatial, and regulatory evidence addresses the fact that protein assemblies form under specific cellular states. HGST uses multi-source biological knowledge to construct spatiotemporal subnetworks and hypergraphs \cite{r28}. DHPRL models dynamic heterogeneous protein information networks \cite{r56}; MComplex combines time-series expression, dynamic PPI construction, graph-learning components, and Mapper analysis \cite{r57}; and STRPCI integrates RNA-protein interactions, time-course expression, localization, and PPI topology through spatiotemporal heterogeneous representation learning \cite{r58}. These methods are biologically richer but require careful alignment of data sources and stronger independent replication.

\section{Computational Integration Strategies}
\begin{figure}[!t]
\centering
\includegraphics[width=\textwidth]{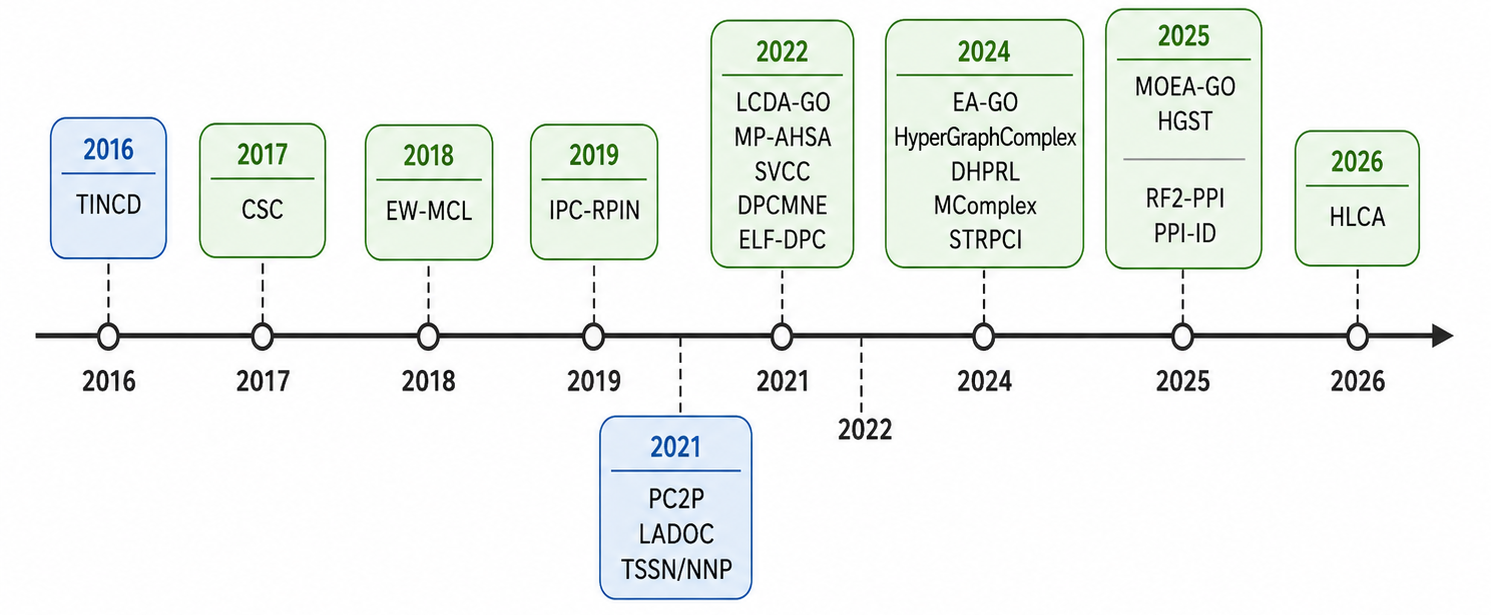}
\caption{Timeline of selected direct and adjacent methods. Method labels are grouped by year to show the transition from topology-centered clustering toward evidence-aware, dynamic, heterogeneous, and hypergraph-oriented approaches. RF2-PPI and PPI-ID are shown separately within 2025 as adjacent PPI/interface context rather than direct complex-detection benchmarks, and ECHO-PPI is omitted from the main timeline and discussed only as preprint context. \emph{Source note: years and method labels were synthesized from cited source papers.}}
\label{fig:timeline}
\end{figure}

\begin{figure}[!t]
\centering
\includegraphics[width=\textwidth]{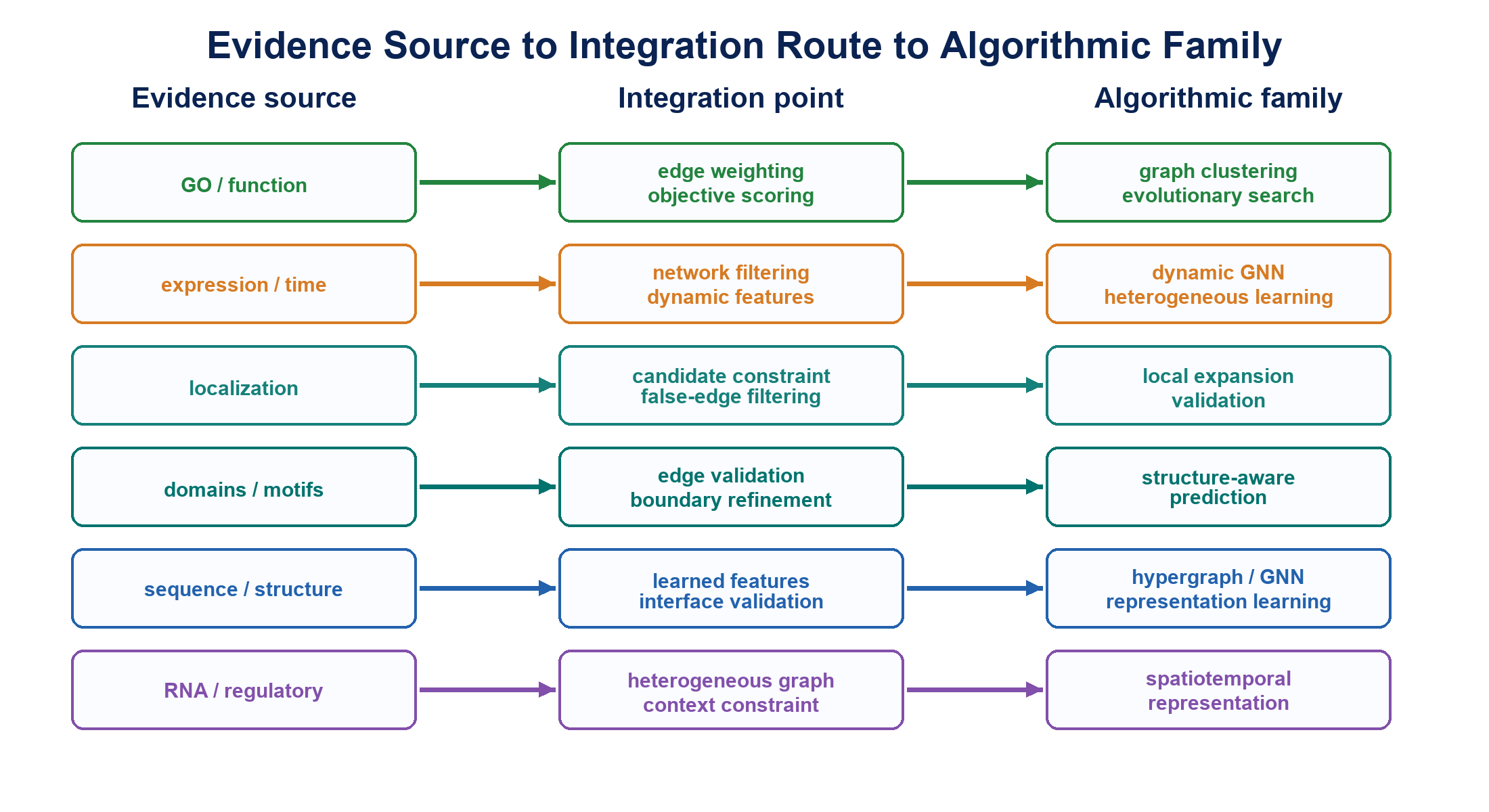}
\caption{Route map for biological evidence integration. The figure links evidence sources to integration points and algorithmic families, emphasizing that the same data type can support filtering, reweighting, learned representation, structural validation, or dynamic heterogeneous modeling. \emph{Source note: conceptual taxonomy derived from the reviewed methods.}}
\label{fig:taxonomy}
\end{figure}

Figure~\ref{fig:timeline} summarizes selected direct and adjacent methods using status categories, while Fig.~\ref{fig:taxonomy} organizes evidence sources, integration mechanisms, and algorithmic paradigms.

Topology-driven methods include density, flow, clique, overlap, cost-based, and parameter-free network formulations \cite{r7,r8,r9,r10,r11,r34}. Edge reweighting and network filtering incorporate GO, expression, localization, or domain evidence before clustering \cite{r15,r18,r19,r23,r25,r29,r30,r31}. Seed selection and local expansion methods use local topology and biological coherence to grow candidates \cite{r17,r20,r21,r22}. Objective-function and evolutionary approaches optimize topology together with biological coherence or multi-property scores \cite{r23,r26,r27}. Supervised, ensemble, and embedding methods learn representations or decision rules from candidate complexes \cite{r24,r33,r51,r60}. Hypergraph, GNN, dynamic, and heterogeneous methods model higher-order and context-dependent relationships beyond static pairwise edges \cite{r28,r52,r56,r57,r58,r59}.

Table~\ref{tab:methods} classifies the reviewed methods by direct complex-detection status, evidence use, organism or benchmark context, overlap handling, and reproducibility risk.

\begingroup
\footnotesize
\setlength{\tabcolsep}{2pt}
\renewcommand{\arraystretch}{1.08}
\begin{longtable}{p{0.12\textwidth}p{0.12\textwidth}p{0.25\textwidth}p{0.20\textwidth}p{0.18\textwidth}}
\caption{Direct Complex-Detection Methods and Evidence Characteristics}\label{tab:methods}\\
\toprule
Method & Status & Main evidence & Structural/interface relevance & Reproducibility risk \\
\midrule
\endfirsthead
\caption[]{Direct Complex-Detection Methods and Evidence Characteristics (continued)}\\
\toprule
Method & Status & Main evidence & Structural/interface relevance & Reproducibility risk \\
\midrule
\endhead
CSC (2017) \cite{r17} & Direct & Topology and GO & Indirect functional support & Annotation dependence \\
TINCD (2016) \cite{r53} & Direct & PPI clustering and TAP/co-complex evidence & Co-complex evidence, not interface evidence & Heterogeneous input dependence \\
EW-MCL (2018) \cite{r18} & Direct & PPI topology and expression & Indirect condition support & Expression-context mismatch \\
IPC-RPIN (2019) \cite{r19} & Direct & PPI, GO, and expression & Functional rather than interface evidence & Threshold sensitivity \\
PC2P (2021) \cite{r34} & Direct & Topology-derived node affinity & Biological evidence is indirect & Limited biological annotation \\
LADOC (2021) \cite{r20} & Direct & Link structure with GO-supported evaluation & Indirect overlap support & Link-graph cost \\
TSSN/NNP and LCDA-GO (2021-2022) \cite{r21,r22} & Direct & PPI and GO & Functional coherence & Seed or local-rule sensitivity \\
MP-AHSA, EA-GO, and MOEA-GO (2022-2025) \cite{r23,r26,r27} & Direct & GO, expression, localization, and topology & Localization supports plausibility & Search parameters and objectives \\
SVCC, DPCMNE, ELF-DPC, and GHAE (2022-2023) \cite{r24,r33,r51,r60} & Direct & Supervised, ensemble, or embedding features & Learned evidence can be difficult to interpret & Training leakage and benchmark separation \\
NRAGE-WPN (2022) \cite{r25} & Direct & Expression and weighted topology & Condition support is indirect & Expression-context mismatch \\
HyperGraph\newline Complex (2024) \cite{r52} & Direct & PPI and sequence in hypergraph learning & Sequence-derived evidence supports protein-centric modeling & Limited independent replication \\
HGST, DHPRL, MComplex, STRPCI, and HLCA (2024-2026) \cite{r28,r56,r57,r58,r59} & Direct; HLCA very recent & Temporal, heterogeneous, RNA-protein, localization, sequence, and higher-order topology & Stronger context and structural plausibility, but variable interface specificity & Multi-source alignment and benchmark harmonization \\
\bottomrule
\end{longtable}
\endgroup

\section{Evaluation Datasets, Metrics, and Benchmark Comparability}
\begin{figure}[!t]
\centering
\includegraphics[width=\textwidth]{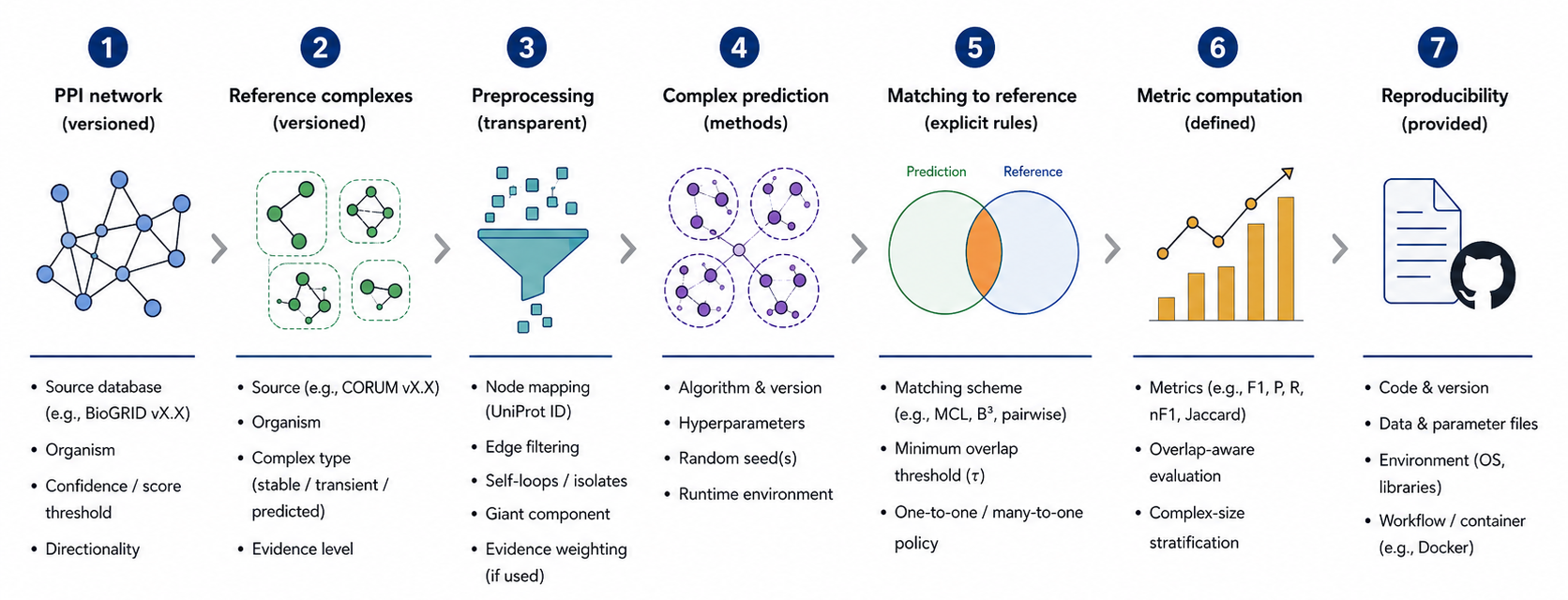}
\caption{Evaluation workflow for comparable benchmarking. The workflow emphasizes that direct comparison requires versioned PPI networks and reference complexes, transparent preprocessing, specified prediction methods, explicit matching rules, defined metric computation, and a reproducibility package containing code, data, parameters, and environment information. Reported performance values are directly comparable only when these protocol components are aligned. \emph{Source note: conceptual evaluation workflow synthesized from reviewed benchmarking practices.}}
\label{fig:evaluation-workflow}
\end{figure}

Fair comparison requires a fixed PPI release, a fixed reference-complex set, shared preprocessing, identical matching rules, identical metric implementations, and an executable software environment. Figure~\ref{fig:evaluation-workflow} summarizes this requirement, and Table~\ref{tab:datasets} summarizes common resources that appear in source-specific evaluation protocols.

\begin{table}[!t]
\caption{Common Datasets and Benchmark Resources}
\label{tab:datasets}
\centering
\footnotesize
\begin{tabularx}{\textwidth}{p{0.15\textwidth}p{0.28\textwidth}Y}
\toprule
Resource & Typical role & Main limitation \\
\midrule
DIP \cite{r35} & Curated PPI network input & Release/version changes affect comparability \\
HPRD \cite{r36,r43} & Human PPI and annotation resource & Legacy database; update status requires caution \\
STRING \cite{r41} & Integrated interaction evidence & Mixes physical and functional associations \\
MIPS \cite{r37} & Yeast reference complexes and annotations & Legacy coverage and versioning issues \\
CYC2008 \cite{r38} & Yeast complex benchmark & Yeast-specific and not directly transferable to human \\
PCDq \cite{r39} & Human complex reference with quality index & Coverage and update status require caution \\
CORUM \cite{r40} & Curated mammalian complexes & Versioning and curation filters affect evaluation \\
\bottomrule
\end{tabularx}
\end{table}

\begin{figure}[!t]
\centering
\includegraphics[width=\textwidth]{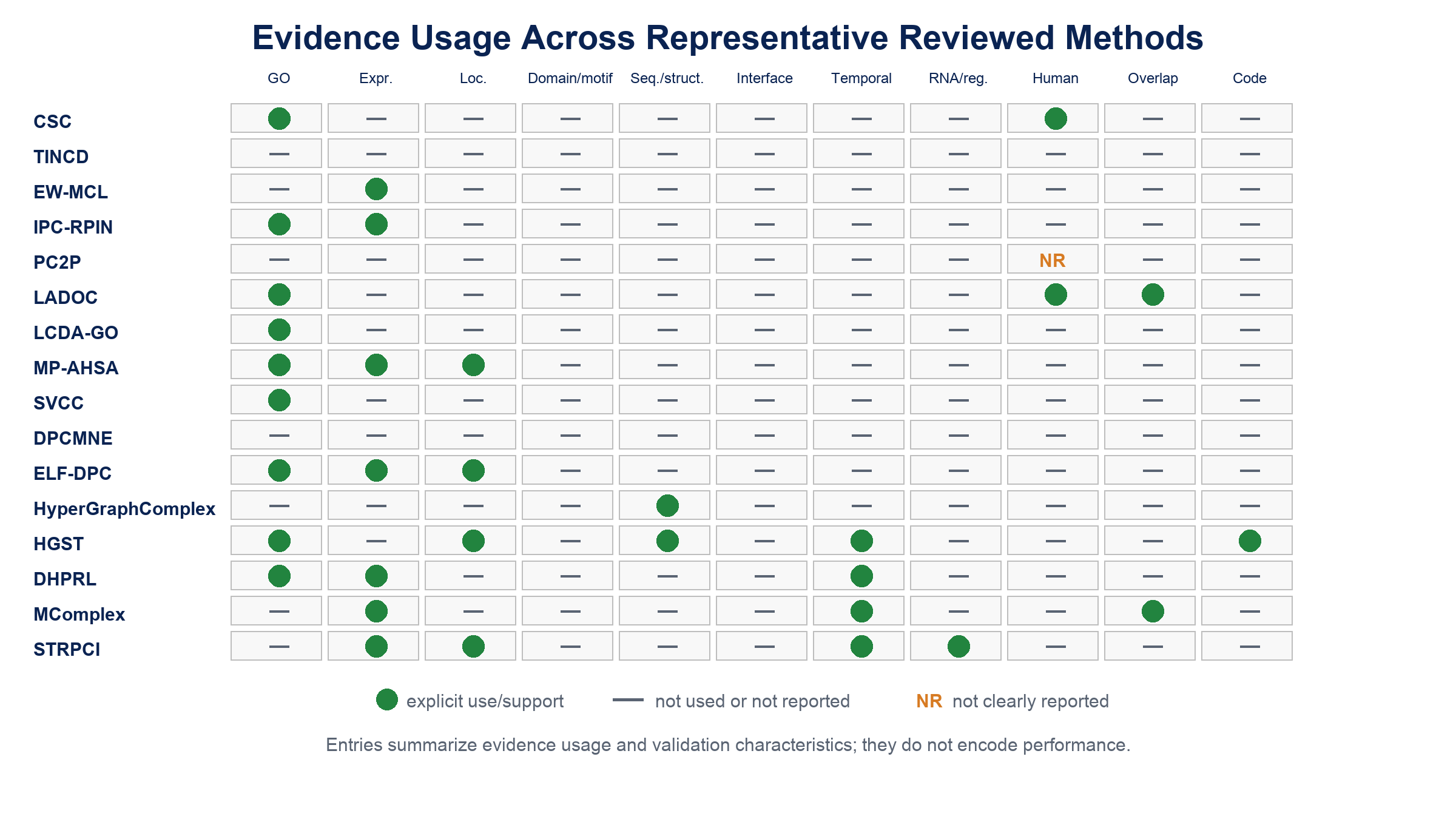}
\caption{Evidence usage across representative reviewed methods. Filled circles indicate explicit use or support in source studies; dashes indicate not used or not reported; NR indicates support was not clearly reported. The matrix summarizes evidence usage and validation characteristics and does not encode performance. \emph{Source note: qualitative matrix synthesized from cited source papers and method descriptions.}}
\label{fig:evidence-matrix}
\end{figure}

Precision, recall, F-measure, sensitivity, PPV, MMR, Rand index, adjusted Rand index, and NMI capture different aspects of complex matching and clustering agreement \cite{r13,r14,r44,r45,r46}. For overlapping complexes, partition-based metrics can understate biologically valid multiple membership unless adapted. Algorithmic reproducibility requires code and environment control; benchmark comparability requires fixed datasets and metrics; biological validity requires evidence that predicted groups correspond to plausible molecular assemblies. Fig.~\ref{fig:evidence-matrix} summarizes evidence usage, while Table~\ref{tab:performance} records protocol dependencies that determine whether source-specific reports could be replicated under a unified benchmark.

\begin{table}[!t]
\caption{Protocol Dependencies for Source-Specific Performance Reports}
\label{tab:performance}
\centering
\scriptsize
\begin{tabularx}{\textwidth}{p{0.13\textwidth}p{0.17\textwidth}p{0.17\textwidth}p{0.16\textwidth}p{0.10\textwidth}Y}
\toprule
Method & PPI context & Reference context & Metric dependency & Code & Unified replication? \\
\midrule
CSC \cite{r17} & DIP/HPRD protocols & MIPS, CYC, PCDq & Preprocessing dependent & Source dependent & Partial \\
IPC-RPIN \cite{r19} & Yeast PPI & MIPS/CYC-type & Threshold dependent & Source dependent & Partial \\
LADOC \cite{r20} & Yeast/human PPI & CYC, CORUM, PCDq & Overlap protocol & Source dependent & Partial \\
LCDA-GO \cite{r22} & Krogan-type PPI & Source complexes & Local matching rules & Source dependent & Partial \\
MP-AHSA \cite{r23} & Multiple yeast PPIs & Standard yeast sets & Parameter protocol & Source dependent & Partial \\
SVCC \cite{r24} & DIP and other PPIs & Source complexes & Train/test split & Source dependent & Partial \\
NRAGE-WPN \cite{r25} & Yeast/human PPI & Source complexes & Metric implementation & Source dependent & Partial \\
EA-GO and MOEA-GO \cite{r26,r27} & Yeast PPI & Source yeast sets & GO/objective dependent & Source dependent & Partial \\
\bottomrule
\end{tabularx}
\vspace{0.3em}
\footnotesize This table records why source-specific performance reports cannot be interpreted as direct rankings. Numerical F-measures are provided only in the supplementary information and should be used only after verification against original datasets, thresholds, and metric implementations.
\end{table}

\section{Critical Comparison of Method Families}
Across the reviewed literature, the strongest current tradeoff for routine use is offered by transparent evidence-aware graph methods that combine topology with interpretable biological signals such as GO, expression, localization, and domain or interface evidence. This is not because they always report the largest source-specific F-measures, but because their evidence pathways can be inspected, ablated, and reproduced more easily than many deep, dynamic, or hypergraph pipelines. Structure-aware, temporal, heterogeneous, and hypergraph methods are biologically attractive because they better match the molecular reality of assembly, but their advantage will remain uncertain until evaluated under shared benchmarks.

Table~\ref{tab:critical-appraisal} provides a qualitative appraisal of method families; the labels are not performance rankings.

\begin{table}[!t]
\caption{Critical Appraisal of Method Families. Ratings represent the authors' qualitative synthesis based on the transparency of evidence pathways, availability of implementation details, number of aligned data sources, and extent of independent validation; they are not quantitative scores.}
\label{tab:critical-appraisal}
\centering
\footnotesize
\begin{tabularx}{\textwidth}{p{0.20\textwidth}p{0.14\textwidth}p{0.18\textwidth}p{0.16\textwidth}Y}
\toprule
Method family & Biological realism & Molecular interpretability & Reproducibility & Main risk \\
\midrule
Topology-only clustering & Low-medium & Medium & High & Confuses graph density with assembly \\
GO-integrated graph methods & Medium-high & High & Medium & GO circularity and annotation bias \\
Expression-aware and dynamic methods & High & Medium & Medium & Condition mismatch and sparse time series \\
Localization, domain, and interface-aware methods & High & High & Medium & Incomplete or coarse annotations \\
Supervised, ensemble, and embedding methods & Medium-high & Mixed & Medium-low & Benchmark leakage and limited interpretability \\
Hypergraph and heterogeneous graph methods & High & Mixed & Low-medium & Complex data alignment and limited replication \\
Foundation-model-assisted approaches & Emerging & Mixed & Low & Overclaiming without complex-level validation \\
Evidence-bundled and uncertainty-aware approaches & High & High & Emerging & Calibration and benchmark standardization \\
\bottomrule
\end{tabularx}
\end{table}

\section{Reproducibility, Annotation Leakage, and Transferability}
GO-based methods do not necessarily contain leakage, but leakage becomes possible when the same annotation source influences prediction features and benchmark interpretation. Recommended safeguards include GO ablation, randomized or permuted GO controls, held-out annotation policies, time-stamped GO releases, and reporting separate results for sparsely annotated proteins. Similarly, supervised and embedding methods should report whether training labels, candidate generation, and benchmark complexes share hidden evidence sources.

Human transfer remains difficult. Yeast benchmarks are mature, but human interactomes are incomplete, tissue-specific, disease-dependent, and unevenly curated \cite{r36,r39,r40,r43}. Orthology can support cross-species transfer, but conserved proteins do not guarantee conserved assembly context, stoichiometry, or interaction interfaces. Stable core complexes may transfer more reliably than transient or condition-specific assemblies.

Table~\ref{tab:reporting-checklist} lists minimum reporting items for reproducible, overlap-aware, and biologically interpretable protein-complex detection studies.

\begin{table}[!t]
\caption{Minimum Reporting Checklist for Protein Complex Detection Studies}
\label{tab:reporting-checklist}
\centering
\scriptsize
\begin{tabularx}{\textwidth}{p{0.25\textwidth}p{0.33\textwidth}Y}
\toprule
Reporting item & Minimum information & Rationale \\
\midrule
PPI database and version & Database name, release, organism, evidence filters & Fixes graph search space \\
Reference-complex version & Benchmark source, release, complex-size filters & Fixes validation target \\
Preprocessing & Removed nodes/edges, duplicate policy, component handling & Prevents hidden protocol drift \\
Matching rules & Threshold, overlap policy, candidate-reference matching & Defines true positives \\
Metric implementation & Exact formulas and software source & Avoids incompatible metric variants \\
GO controls & GO release, ablation, held-out or permutation policy & Reduces circularity risk \\
Randomness & Seeds, repeated runs, parameter settings & Supports stochastic reproducibility \\
Software environment & Code, dependencies, container, runtime, hardware & Enables independent reruns \\
Structural provenance & Model source, experimental structure source, or prediction method and confidence & Supports structure-aware interpretation \\
Direct/indirect evidence & Separate binary binding, co-complex, functional, and predicted edges & Prevents evidence-type conflation \\
Uncertainty & Confidence intervals, repeated-run variation, calibration & Supports validation prioritization \\
Human validation & Human reference set, tissue/cell context, or absence statement & Clarifies transferability \\
\bottomrule
\end{tabularx}
\end{table}

\section{Research Priorities}
\begin{figure}[!t]
\centering
\includegraphics[width=\textwidth]{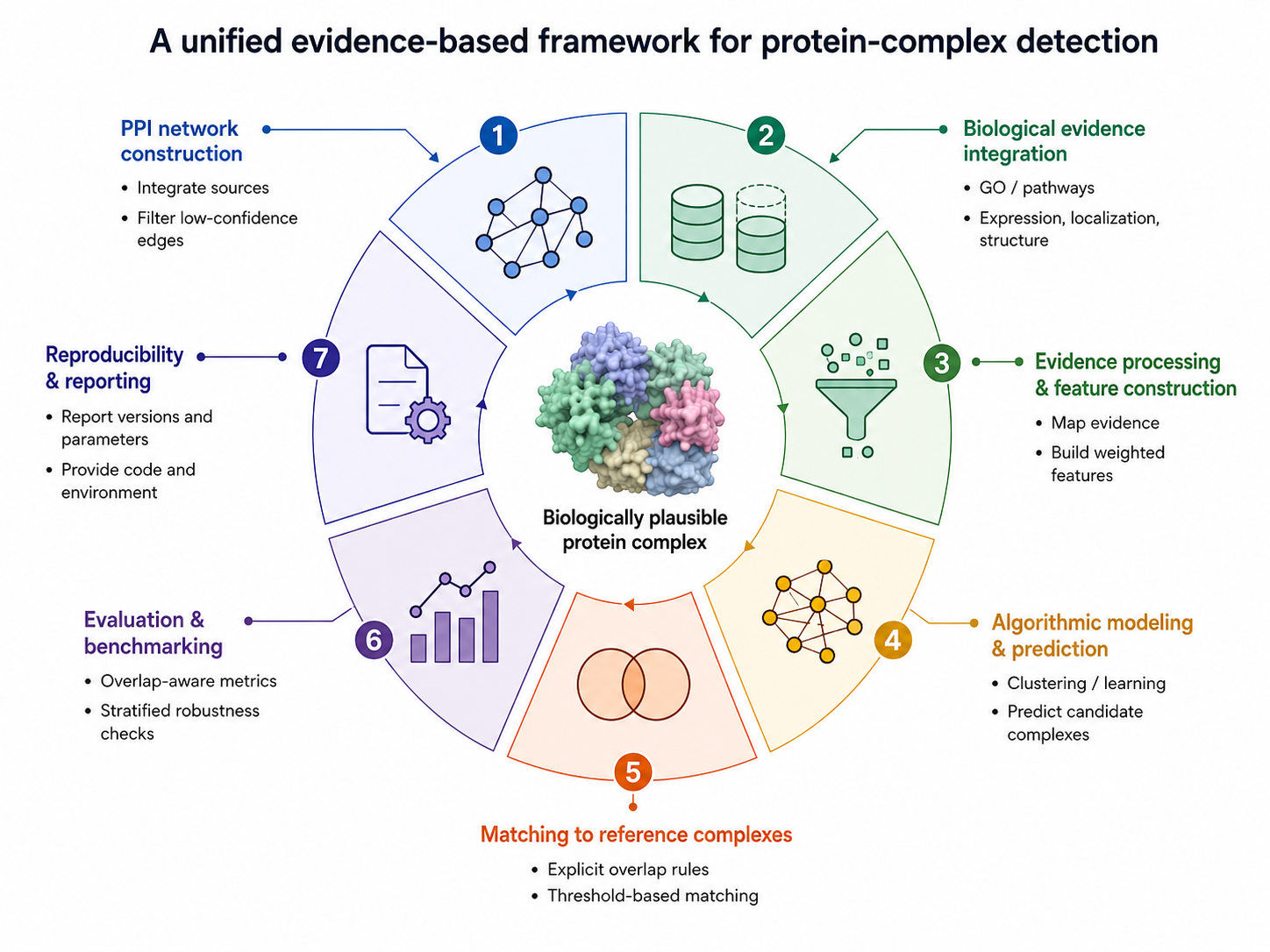}
\caption{Unified evidence-based framework for protein-complex detection. The framework links PPI network construction, biological evidence integration, evidence processing and feature construction, algorithmic modeling and prediction, matching to reference complexes, evaluation and benchmarking, and reproducibility reporting around the goal of biologically plausible protein-complex prediction. \emph{Source note: conceptual framework synthesized from reviewed methods and limitations.}}
\label{fig:future-directions}
\end{figure}

Figure~\ref{fig:future-directions} summarizes a unified evidence-based framework for the next generation of protein-centric complex-detection methods.

Priority 1 is unified, versioned, executable benchmarking. The biological motivation is simple: without common inputs and metrics, reported differences cannot be mapped to biological improvement. The computational requirement is a benchmark bundle with PPI releases, reference complexes, preprocessing scripts, matching rules, metric code, seeds, and containers.

Priority 2 is human, tissue-specific, and condition-specific validation. Human assemblies are influenced by tissue, disease state, cell type, and curation bias. Near-term progress requires versioned human PPI releases, confidence scores, and reference complexes stratified by evidence and context.

Priority 3 is domains, motifs, interfaces, and structure-aware evidence. Mechanistic evidence can filter implausible edges, distinguish direct from indirect association, refine complex boundaries, and prioritize candidates. Evaluation should test whether structure-derived features improve complex-level validity rather than only pairwise PPI prediction.

Priority 4 is temporal, spatial, regulatory, heterogeneous, and higher-order modeling. Dynamic and hypergraph methods better reflect cellular assembly, but they require aligned multi-source inputs and independent benchmark replication.

Priority 5 is uncertainty-aware protein language models and structure-derived representations. These models can supply features for under-studied proteins, but complex-detection claims require calibrated uncertainty and benchmarked assembly-level validation \cite{r32,r49,r50}. ECHO-PPI is relevant only as an emerging 2026 arXiv preprint example of evidence-bundled overlap assignment, not as a peer-reviewed or independently benchmarked method \cite{r55}.

Priority 6 is cross-species transfer and evaluation for under-studied organisms. Transfer learning should report orthology assumptions, conserved-interface evidence, organism-specific annotation gaps, and whether validation targets stable complexes or condition-specific modules.

\section{Practical Recommendations}
For fast baseline comparison, pair a new method with MCL, MCODE, ClusterOne-style overlap detection, and PC2P where feasible \cite{r7,r8,r10,r34}. For functionally coherent yeast complexes, use GO-integrated methods but report GO controls. For condition-specific complexes, use expression-aware or dynamic models only when the expression context matches the biological question. For human discovery, favor methods with explicit human validation and avoid assuming yeast transfer. For mechanistic prioritization, combine graph evidence with domain, motif, interface, sequence, or structural evidence, and report whether the evidence supports direct binding or only co-complex association.

\section{Limitations of This Review}
This article is a focused critical review rather than a systematic meta-analysis. Reporting is inconsistent across source studies, and not all software, benchmark releases, thresholds, or metric implementations are available. Recent deep, hypergraph, dynamic, and heterogeneous methods have limited independent replication. The field remains yeast-biased, overlap-aware metrics are not uniformly applied, and physical complexes are not always clearly separated from functional modules. The 2024-2026 literature is evolving quickly; very recent papers and preprints should be interpreted as emerging evidence rather than settled consensus.

\section{Conclusions}
Topology remains necessary for protein-complex detection, but it is not sufficient. The objective is not merely to find dense graph regions; it is to identify biologically plausible molecular assemblies supported by function, localization, expression, interaction mechanism, sequence or structure, temporal context, and reproducible evaluation. Transparent evidence-aware graph methods currently provide a practical balance of interpretability and reproducibility. Structure-aware, dynamic, heterogeneous, and hypergraph models are promising because they better reflect molecular assembly, but they require shared benchmarks and independent validation before source-specific performance claims can be interpreted as field-wide progress.

Future progress will depend on connecting molecular evidence to executable evaluation. Methods should distinguish direct binding from co-complex association, report calibrated uncertainty, support overlap-aware assessment, archive code and datasets, and make their biological evidence traceable. Protein-complex detection will be most useful for structural and functional bioinformatics when graph predictions can be translated into testable hypotheses about molecular interfaces, cellular context, and experimentally prioritizable assemblies.

\begin{itemize}
\item Protein-complex detection should be interpreted as molecular assembly inference, not only dense-subgraph discovery.
\item Biological evidence from GO, expression, localization, domains, motifs, interfaces, sequence, structure, temporal context, and regulatory data improves plausibility but introduces reproducibility risks.
\item Direct complex-detection methods must be distinguished from adjacent PPI, interface, motif-mapping, and structure-prediction tools.
\item Source-specific F-measures are not rankings unless PPI releases, reference complexes, matching thresholds, preprocessing, metrics, and software environments are fixed.
\item Future methods need structure-aware evidence, overlap-aware evaluation, uncertainty calibration, and executable benchmark bundles.
\end{itemize}

\section*{Author Biographies}
Sima Soltani, Mehrdad Jalali, Yahya Forghani, and Reza Sheibani work on computational biology, artificial intelligence, data science, and network-based analysis of biological systems, with interests in protein interaction networks, evidence integration, and reproducible bioinformatics methods.

\section*{Funding}
The authors received no specific funding for this work.

\section*{Conflict of interest}
The authors declare no competing interests.

\section*{Data availability}
No new datasets were generated for this review. All data discussed in this article are available from the cited sources. The complete extraction matrix, search strings, and source-specific benchmark information are provided in the supplementary information.

\section*{Code availability}
No new protein complex detection algorithm was developed for this review. The figure-generation script used to prepare the conceptual figures is included in the submission package.

\section*{Author contributions}
S.S. contributed to literature collection, method review, and manuscript drafting. M.J. supervised the study, refined the methodological framing, contributed to critical revision, and served as corresponding author. Y.F. contributed to review design, interpretation, and manuscript revision. R.S. contributed to manuscript review, technical checking, and final revision. All authors reviewed and approved the final manuscript.

\section*{Acknowledgements}
No acknowledgements are declared.

\section*{AI-assisted writing disclosure}
During manuscript preparation, AI-assisted tools were used for language polishing, formatting support, and technical consistency checks. The authors reviewed and edited all AI-assisted outputs and take full responsibility for the final content.

\end{document}